\documentclass[aps,prl,reprint,showpacs,superscriptaddress,fleqn]{revtex4-1}

\usepackage{graphicx}
\usepackage[utf8]{inputenc}
\usepackage[english]{babel}
\usepackage{color}
\usepackage{soul}
\usepackage{amsmath}

\newcommand{\VJ}[1]{\textcolor{black}{#1}}

\begin{document}

\title{Strength and length-scale of the interaction between domain walls\\ and pinning disorder in thin ferromagnetic films}


\author{P. G\'ehanne}
\affiliation{Universit\'e Paris-Saclay, CNRS,  Laboratoire de Physique des Solides, 91405, Orsay, France.}
\author{S. Rohart}
\affiliation{Universit\'e Paris-Saclay, CNRS,  Laboratoire de Physique des Solides, 91405, Orsay, France.}
\author{A. Thiaville}
\affiliation{Universit\'e Paris-Saclay, CNRS,  Laboratoire de Physique des Solides, 91405, Orsay, France.}
\author{V. Jeudy}
\email{vincent.jeudy@universite-paris-saclay.fr}
\affiliation{Universit\'e Paris-Saclay, CNRS,  Laboratoire de Physique des Solides, 91405, Orsay, France.}

\date{\today}
\begin{abstract}
We explore the magnetic-field-driven motion of domain walls with different chiralities in thin ferromagnetic films made of Pt/Co/Pt, Au/Co/Pt, and Pt/Co/Au. From the analysis of domain wall dynamics, we extract parameters characterizing the interaction between domain walls and weak 
pinning disorder of the films.
The variations of domain wall structure, controlled by an in-plane field, are found to modify the characteristic length-scale of pinning in strong correlation with the domain wall width, whatever its chirality and the interaction strength between domain walls and pinning defects. These findings should be also relevant for a wide variety of elastic interfaces moving in weak pinning disordered media. 

\end{abstract}

%

\maketitle

The controlled motion of magnetic textures such as chiral domain walls (DWs)~\cite{thiaville_EPL_2012}
or skyrmions resulting from the Dzyaloshinskii-Moriya interaction (DMI)~\cite{caretta_natnano_2018}  
are at the basis of potential applications of spintronic devices~\cite{fert_natnano_2013}. 
However, magnetic textures are very sensitive to weak pinning due to ubiquitous inhomogeneities of magnetic materials, which strongly reduces their velocity and produces stochastic universal behaviors~\cite{ferrero_prl_2017_spatiotemporal_patterns,grassi_prb_2018}.
Despite numerous recent studies~\cite{je_prb_2013_DMI,lavrijsen_prb_2015,lau_prb_2016,pellegrin_prl_2017,lau_prb_2018,hartmann_prb_2019} focusing on the dynamics of pinned chiral magnetic textures, their
interactions with weak pinning disorder is far from being understood.



DWs are well known to present universal behaviors~\cite{lemerle_PRL_1998_domainwall_creep,jeudy_PRL_2016_energy_barrier,diaz_PRB_2017_depinning} similar to those encountered by interfaces in a wide variety of other physical systems. Those behaviors can be described by minimal statistical-physics models~\cite{edwards_wilkinson_1982,kolton_prb_2009_pathways,agoritsas_physicaB_12} as an interplay between interface elasticity, weak pinning, thermal activation, and a driving force $f$. The dynamical regimes of interfaces strongly depend on the relative magnitude of the drive $f$ compared to a depinning threshold force $f_d$. 
In the creep regime ($f<f_d$), the velocity follows an Arrhenius law $v \sim e^{-\Delta E/(k_BT)}$, where the barrier height presents an asymptotic power law behavior $\Delta E \sim f^{-\mu}$ with the force $f$ close to zero. At and just above the depinning threshold, the velocity presents a power law variation with the temperature $v(f=f_d) \sim T^\psi$ and drive $v(f \gtrsim f_d) \sim (f-f_d)^\beta$, respectively. The critical exponents $\mu$, $\psi$, and $\beta$ are universal (i.e. material and temperature independent). Their values characterize the universality class of the motion and reflects the dimension of the interface and embedding medium, the range of elasticity and the interaction with pinning defects.
In ultrathin films with perpendicular anisotropy, a perpendicular magnetic field $H$ can serve as an isotropic driving force ($H \propto f$). A large majority of experimental studies on DW dynamics reported in the literature~\cite{jeudy_PRB_2018_DW_pinning} is compatible with the theoretical predictions ($\mu=1/4$, $\psi=0.15$, and $\beta=0.25$) for short range (random bond) interactions between pinning disorder and DWs.
However, the minimal models 
ignore the exact structure of interfaces and the characteristic length-scale of pinning is a parameter chosen arbitrarily~\cite{agoritsas_physicaB_12,agoritsas_PRE_13}. Since the seminal work of Lemerle {\it et al.}~\cite{lemerle_PRL_1998_domainwall_creep} on the creep motion, basic issues such as the length-scale and the strength of interaction between DW and defects in magnetic materials remain open.


Recent experiments on the creep motion of chiral DWs have evidenced the correlations between the DW magnetic texture and its dynamics. 
In thin films with perpendicular anisotropy, the DMI results in an in-plane effective magnetic field $H_{\mathrm{DMI}}$ pointing in the direction perpendicular to the DW. The DMI field combined with an in-plane field $H_x$ can be used to adjust the in-plane component of magnetization direction in the DW and to control the DW magnetic structure. The recent observation of asymmetric expansion of initially circular domains~\cite{je_prb_2013_DMI} has lead to precise investigations of the variation of DW width, energy and stiffness~\cite{lau_prb_2016,pellegrin_prl_2017,lau_prb_2018,hartmann_prb_2019} with the direction and magnitude of the in-plane and DMI fields and to re-examine more generally the creep motion of chiral DWs~\cite{hartmann_prb_2019}. Rather accurate descriptions of the shape of the velocity curves versus in-plane field are now obtained~\cite{lau_prb_2018,hartmann_prb_2019}.
The proposed models discuss the variations of creep barrier height $\Delta E$ with the DW energy~\cite{je_prb_2013_DMI,lavrijsen_prb_2015,lau_prb_2016,pellegrin_prl_2017,lau_prb_2018,hartmann_prb_2019}. Surprisingly, the interaction between the DW and random pinning disorder is assumed to be independent of DW's magnetic structure.

%

Here, we evidence a strong correlation between the variation of DW width controlled by an in-plane field and the characteristic length of pinning, in films with three different chiralities.
%
Our argument is organized as follows. 
We first extract the 
material and in-plane field dependent 
pinning parameters controlling DW dynamics, from the self-consistent analysis 
 proposed in Ref.~\cite{jeudy_PRL_2016_energy_barrier, diaz_PRB_2017_depinning}. We then numerically compute the variation with in-plane and DMI fields of the DW energy and width. The latter are compared to the variation of pinning range and strength deduced from the 
pinning parameters via scaling relations~\cite{jeudy_PRB_2018_DW_pinning}.

\textit{Experimental techniques.} 
The samples are Pt/Co/Pt, Pt/Co/Au, and Au/Co/Pt films (with thicknesses of 5~nm for Pt and Au and 0.9~nm for Co) with perpendicular magnetic anisotropy, which have been grown by e-beam evaporation in ultra high vacuum on Si(001)/SiO$_2$(100~nm)/Ta(5~nm) templates. They present (111) oriented crystallites with a typical grain size of 15~nm \cite{gross_PRM_2018}.
The micromagnetic parameters characterizing the films are detailed in Table~\ref{table:table1}. 
\begin{table*}[!htbp]
	\centering
	\begin{ruledtabular}
		\begin{tabular}{l c c c c c c c}
			
			Material &$M_s$ (MA/m) & $\mu_0 H_{K_0}$ (mT) & $\mu_0 H_{DMI}$  (mT) & $K_0$ (kJ/m$^3$) &  $   \Delta_0$ (nm) &$\sigma_0$ (mJ/m$^2$) & $D$ (mJ/m$^2$)\\
			
			\hline
			Pt/Co/Pt 
			& 1.62(0.04) & 580(10)&  0(10) & 470(14) & 5.8(0.1)& 11.0(0.2) & 0(0.10)\\
			Pt/Co/Au
			& 1.65(0.01) & 770(30) & -105(10) & 635(25) & 5.0(0.1)& 12.7(0.3) & -0.87(0.09)\\
			Au/Co/Pt
			& 1.61(0.02) & 900(100) & +78(10) & 726(80) & 4.7(0.3)& 13.6(0.8) & 0.59(0.08)\\
			
		\end{tabular}
		\caption{\label{table:table1}  \textbf{Micromagnetic parameters.}
			For each material, the table indicates the magnetization saturation $M_s$, the anisotropy field $\mu_0 H_{K_0}$, and the DMI field $\mu_0 H_{DMI}$ deduced from SQUID, Kerr anisometry, and velocity measurements, respectively. The effective anisotropy constant $K_0$ is calculated from $K_0=\mu_0 H_{K_0}M_s/2$. The values of the DW width parameter ($\Delta_0=\sqrt{A/K_0}$) and the DW Bloch energy ($\sigma_0 = 4A/\Delta_0$) are obtained assuming a stiffness constant $A=16~\mathrm{pJ/m}$. The interface DMI energy $D$ is calculated from $D=\mu_0 H_{DMI}M_s\Delta_0$.
			The numbers in parenthesis are the error bars.} 
	\end{ruledtabular}
\end{table*}
\begin{figure*}[!tbp]
	\includegraphics[width=8.7 cm]{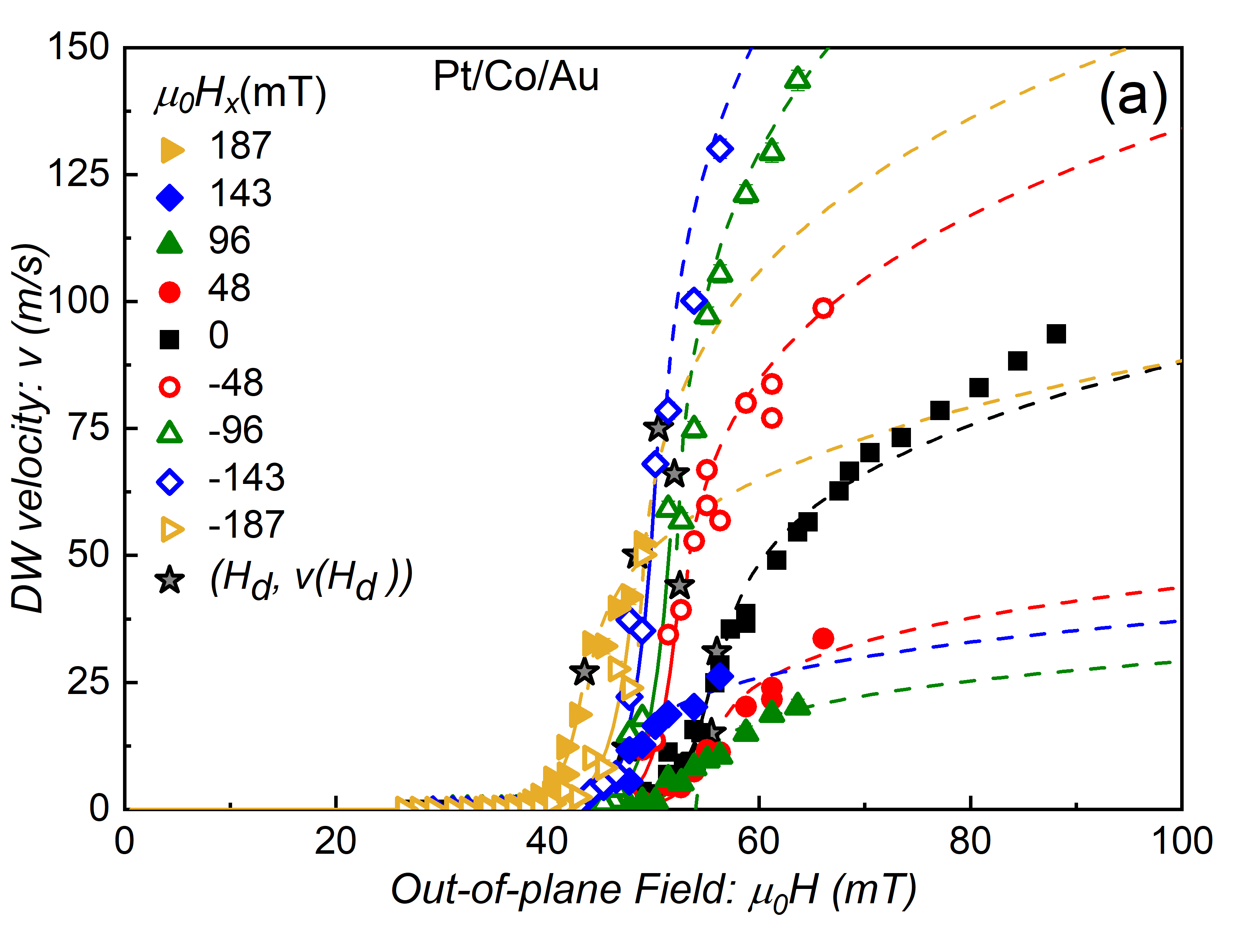}
	\includegraphics[width=8.7 cm]{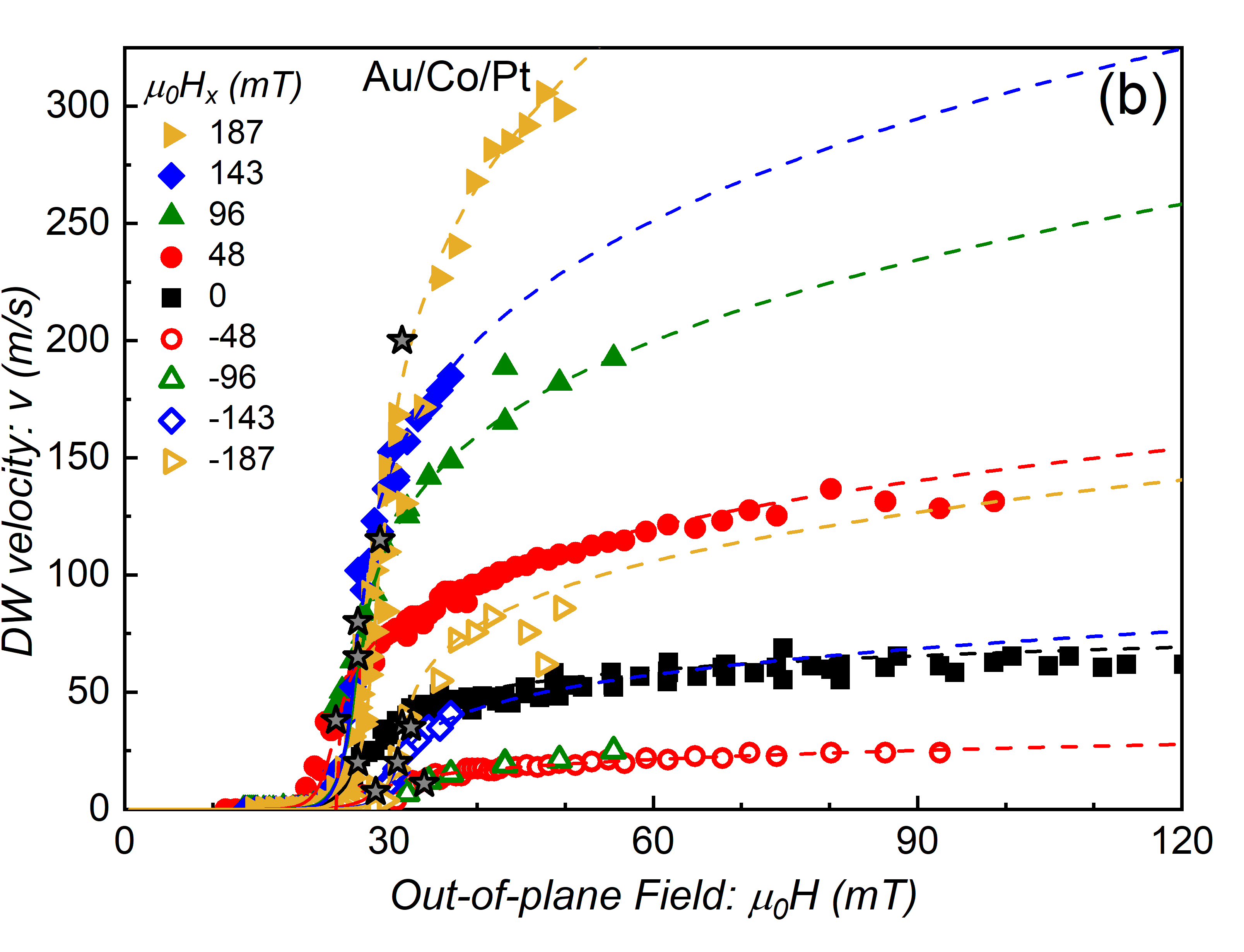}
	
	\includegraphics[width=8.7cm]{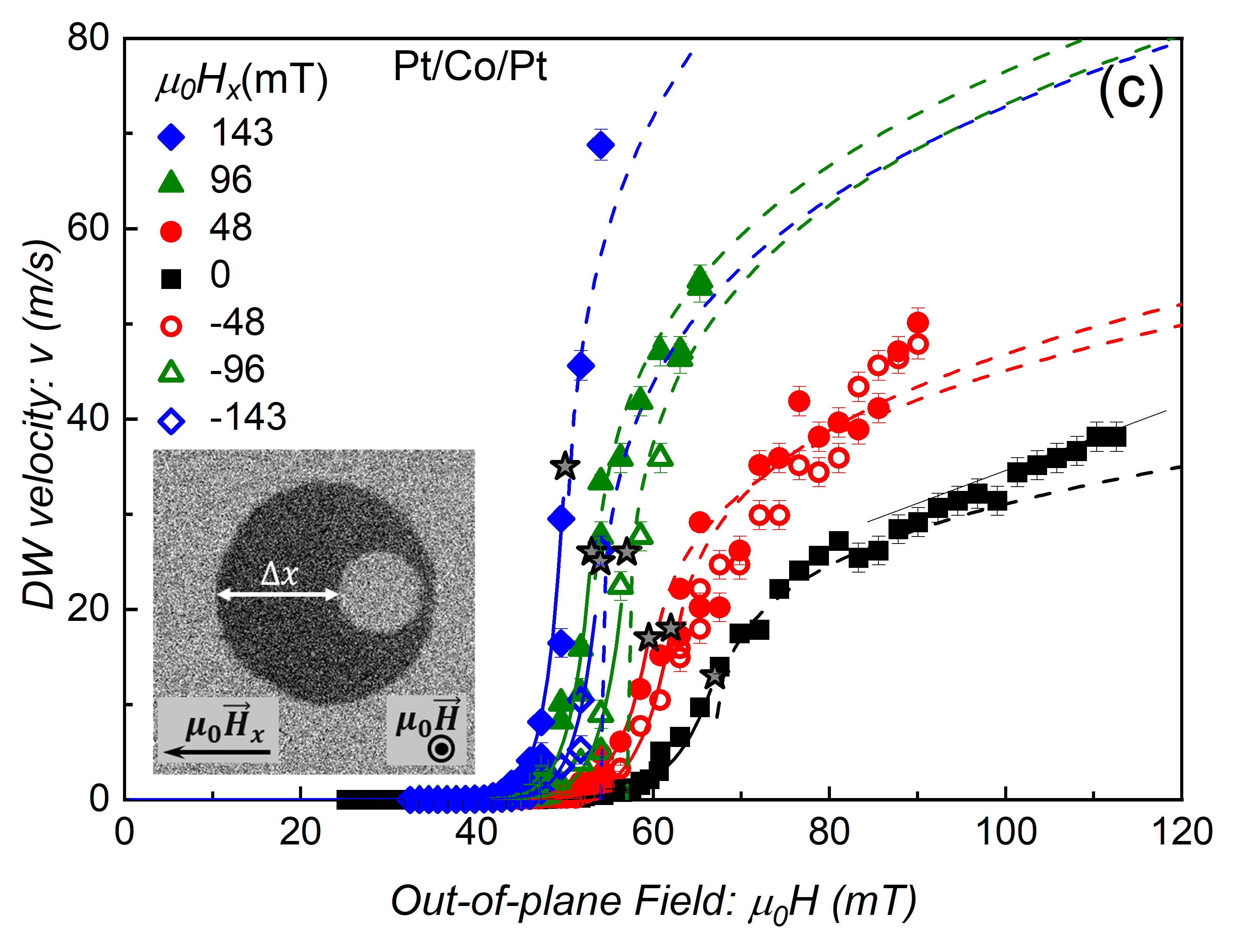}	
	\includegraphics[width=4.1 cm]{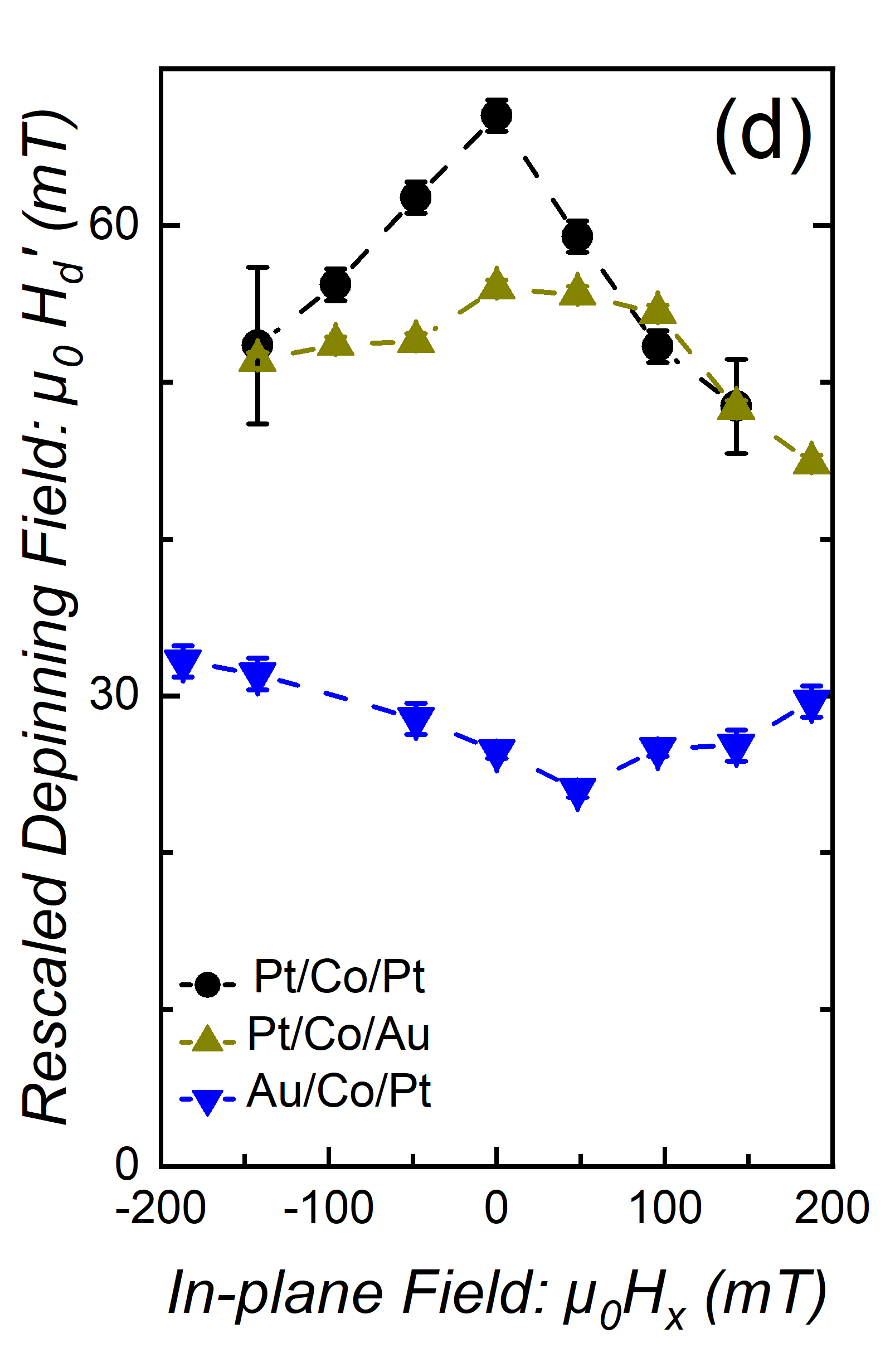}	
	\includegraphics[width=4.1 cm]{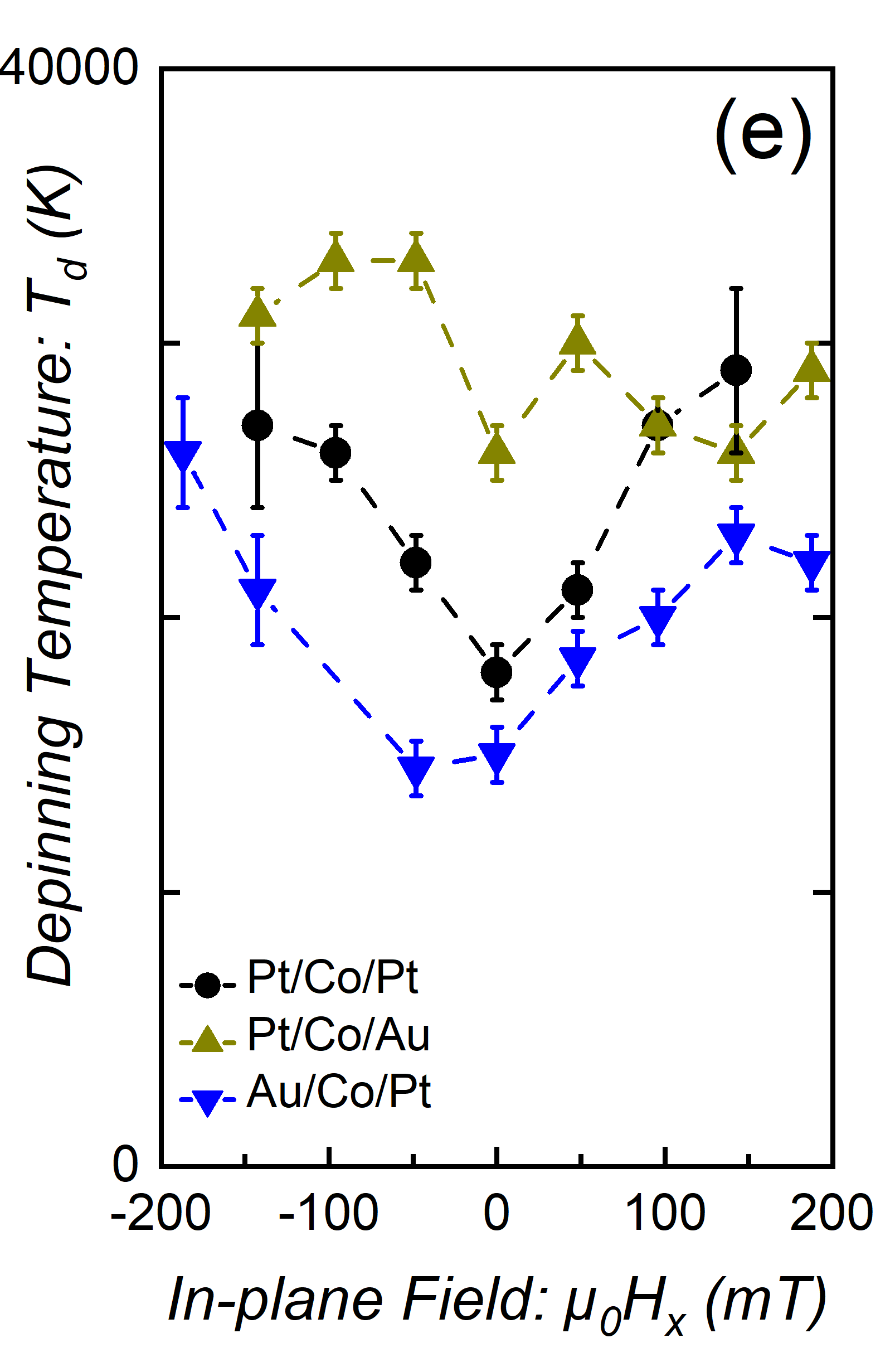}
	\caption{\textbf{Domain wall dynamics}. DW velocity versus out-of-plane magnetic field $\mu_0 H$ measured for the (a) Pt/Co/Au, (b) Au/Co/Pt, and (c) Pt/Co/Pt films and for different values of the in-plane field $\mu_0 H_x$. The solid and dash lines are predictions for the creep and depinning regimes (see Eqs. 1). The stars correspond to the coordinates of the depinning thresholds $(H_d,v(H_d)$). A good agreement with the data is obtained up to the cross-over between the depinning and flow regimes~\cite{diaz_PRB_2017_depinning}, which can be observed at $\mu_0 H_x = 0~\mathrm{mT}$ for Pt/Co/Au ($\mu_0 H> 70~\mathrm{mT}$) and Pt/Co/Pt ($\mu_0 H> 90~\mathrm{mT}$). The curves (d) and (e) correspond to the variation with $\mu_0 H_x$ of the \VJ{rescaled depinning field ($H_d'$, see the text)} and temperature ($T_d$) deduced from the fit of Eqs. \ref{eq: 1}. Insert of c. Displacement of DW in the Au/Co/Pt film produced by a magnetic field pulse ($\mu_0 H=3~\mathrm{mT}$, $\Delta t=1 ~\mathrm{\mu s}$) for $\mu_0 H_x= 48~\mathrm{mT}$. 
	} 
	\label{fig:1}
\end{figure*}
%
The DW displacement was observed by polar Kerr microscopy. The in-plane magnetic field $H_x$ controlling the magnetization direction in the DW was generated by two large coils supplied with DC current.
The out-of-plane field $H$ used to move DW was produced by a small coil (diameter $\approx 1$~mm) placed in the close vicinity of the films and supplied by pulses of duration $\Delta t$ between $1~\mathrm{\mu s}$ and $10~\mathrm{ms}$. The explored range of $\mu_0H_x (\pm 187~\mathrm{mT})$ and $\mu_0 H (0-120~\mathrm{mT})$ was limited by the nucleation of multiple domains, which impede measurement of DW displacement.
The DW velocity $v$ corresponds to the ratio $\Delta x/\Delta t$, where $\Delta x$ was exclusively measured in the direction of $H_x$ (see the inset of Fig.~\ref{fig:1}).

\textit{Domain wall dynamics.}
The velocity curves obtained for the three films and different values of $\mu_0H_x$ are reported in Fig. \ref{fig:1}. 
For their analysis, we used the self-consistent description of the creep and depinning regimes~\cite{jeudy_PRL_2016_energy_barrier,diaz_PRB_2017_depinning,jeudy_PRB_2018_DW_pinning,shahbazi_PRB_2018,shahbazi_PRB_2019}: 
\begin{equation}
v(H)=\left \{
\begin{array}{lr}
v(H_d)\exp (-\frac{\Delta E}{k_B T}) & \mathrm{creep}: H<H_d\\
\frac{v(H_d)}{x_0}(\frac{T_d}{T})^\psi(\frac{H-H_d}{H_d})^\beta & \mathrm{depinning}: H \gtrsim H_d, \\
\end{array}
\right.
\label{eq: 1}
\end{equation}
where $\Delta E=k_B T_d((H/H_d)^{-\mu}-1)$ is the energy barrier of the creep regime and the depinning law is the asymptotic velocity behavior in the limit of negligible contribution of thermal fluctuations~\cite{diaz_PRB_2017_depinning}. In Eq.~\ref{eq: 1},  $\mu=1/4$, $\beta=0.25$, and $\psi=0.15$ are universal critical exponents and $x_0=0.65$ a universal constant~\cite{diaz_PRB_2017_depinning} characterizing the quenched Edwards Wilkinson universality class~\cite{edwards_wilkinson_1982,bustingorry_PRE_12_thermal_rounding, diaz_PRB_2017_depinning} and are therefore fixed. 
The non-universal parameters are the characteristic height of effective pinning barrier $k_B T_d$ and coordinates of the depinning threshold (corresponding to $\Delta E \rightarrow 0$) $H_d$ and $v(H_d)$. Those three parameters depend on the film magnetic and pinning properties, and external parameters as the in-plane field, and the temperature~\cite{jeudy_PRB_2018_DW_pinning}.
%
%
As it can be observed in Fig.~\ref{fig:1} a-c, all the fits of Eqs.~\ref{eq: 1}, obtained with only three fitting parameters, present a good agreement with the velocity curves (see Refs.~\cite{jeudy_PRB_2018_DW_pinning} for details on the fitting procedure). 
This finding is compatible with the results reported for a large variety of other magnetic materials~\cite{jeudy_PRL_2016_energy_barrier,diaz_PRB_2017_depinning,jeudy_PRB_2018_DW_pinning,shahbazi_PRB_2018,shahbazi_PRB_2019} and confirms that chiral and non-chiral DWs follow very similar universal behaviors as predicted in Ref.~\onlinecite{hartmann_prb_2019} for the creep regime. 
\begin{figure*}[ht]
	\includegraphics[width=8.7 cm]{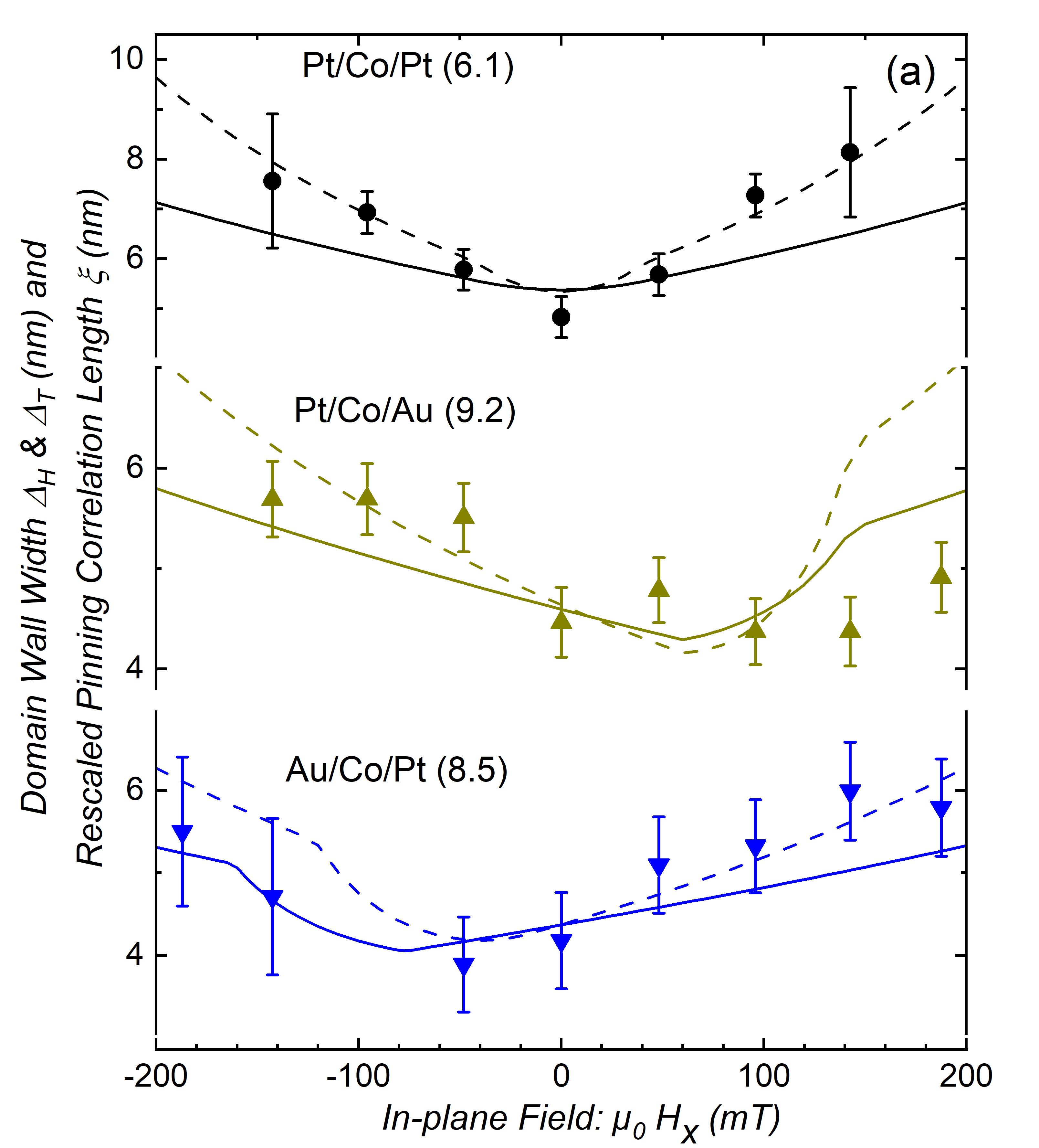}
	\includegraphics[width=8.7 cm]{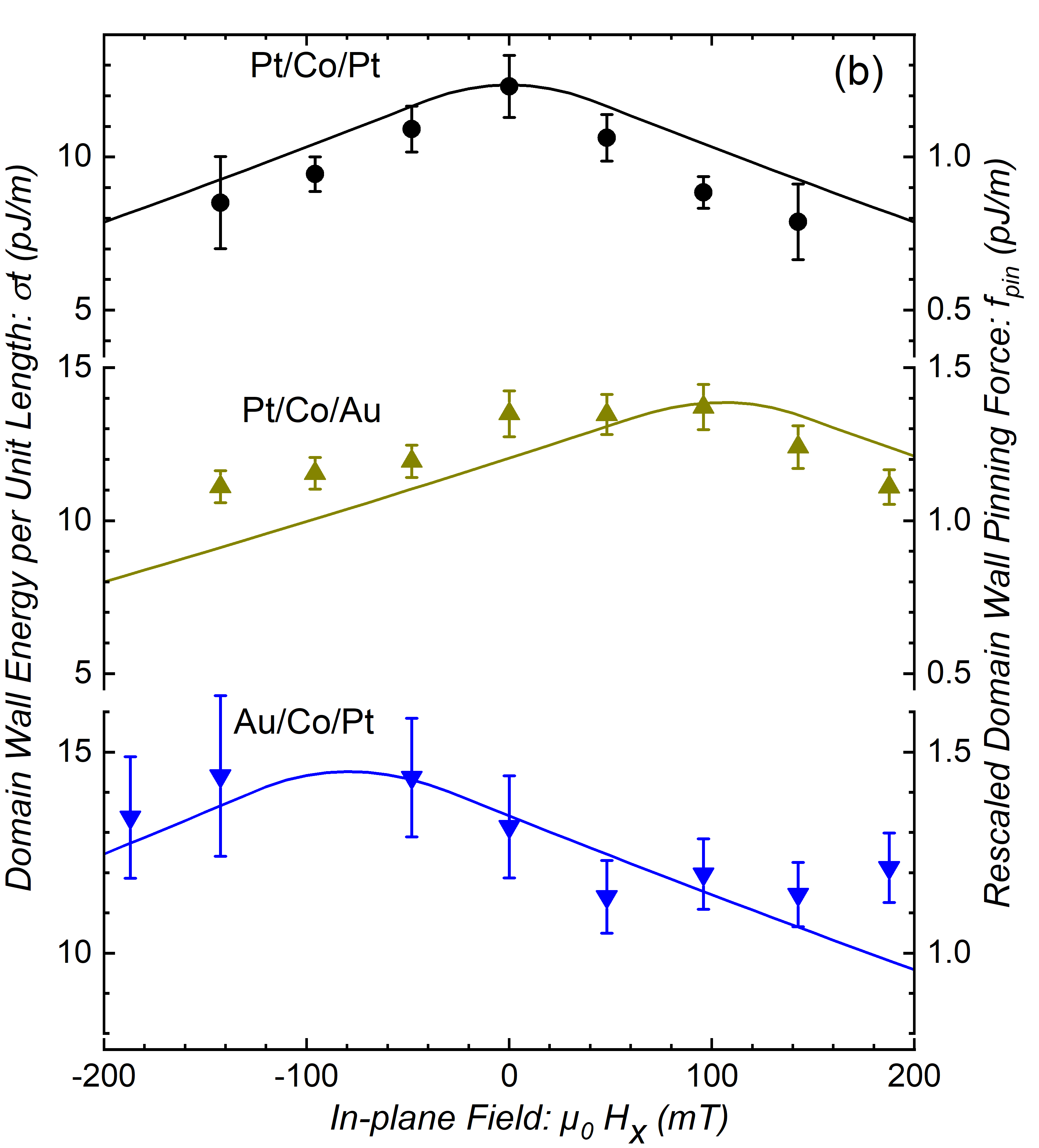}
	
	\caption{\textbf{Pinning of domain wall at microscopic scale}. 
	(a) Domain wall widths $\Delta_H$ (solid line) and $\Delta_T$ (dash line), and rescaled correlation length of the pinning $\xi$ versus in-plane magnetic field $\mu_0 H_x$, for the three films. For each film, the values of $\xi$ deduced from Eq.~\ref{eq: 3_xi} were rescaled by constant ($ \approx \mathrm{min}(\xi)/\mathrm{min}(\Delta_{T})$) indicated in parenthesis to highlight the strong correlation between the DW width and $\xi(\mu_0 H_x)$. 
	(b) Domain wall energy per unit length $\sigma t$ (left vertical scale, solid line) and the rescaled pinning force $f_{pin}$ (right vertical scale) versus $\mu_0 H_x$. For each film, the values of $f_{pin}$ deduced from Eq.~\ref{eq: 2_fpin} were rescaled in order of set $\mathrm{max}(f_{pin}) \approx \mathrm{max}(\sigma t)/10$. 
	} 
	\label{fig:3}
\end{figure*} 
%

\textit{Interaction between Domain wall and defects.}
Having discussed universal behaviors, we can now focus on the obtained non-universal parameters (see Fig.~\ref{fig:1} d-e). (Here, we do not discuss the values of $v(H_d)$ since it is partly determined by the contribution of DW dynamics in the flow regime at the threshold $H_d$~\cite{jeudy_PRB_2018_DW_pinning}.) 
\VJ{
Note that the in-plane field $H_x$ tilts the magnetization in the domains on both sides of the DW, which reduces the driving field to $H'=H\sqrt{1-(H_x/H_{K_0})^2}$, where $H_{K_0}$ is the anisotropy field~\cite{Jue_PRB_2016}. To account for this effect~\cite{note_Hd_effectif}, the depinning field $H_d$ has to be rescaled to $H_d'$ consequently.}

Surprisingly, the values of $H_d'$ and $T_d$ vary with the in-plane magnetic field $H_x$. This strongly suggests that the pinning properties of DWs depend on with their magnetic texture, in contrast to the usual assumption found in the literature~\cite{je_prb_2013_DMI,lau_prb_2016,pellegrin_prl_2017,lau_prb_2018,hartmann_prb_2019}.
%
%
In order to understand the variations of $H_d'$ and $T_d$, we shall discuss the DW pinning at the microscopic scale and, more precisely, to compare the effect of the in-plane field on the DW energy and width and on the strength and length-scale of pinning (see Fig.~\ref{fig:3}). 

To predict accurately the variation of DW energy and structure with in-plane and DMI fields, we have decided not to use simplified descriptions of DWs~\cite{je_prb_2013_DMI,lau_prb_2016,pellegrin_prl_2017,lau_prb_2018,hartmann_prb_2019}. We have calculated the magnetization profile $\vec{M}(x)$ of a DW whose plane is $\perp \vec{H}_x$, from numerical micromagnetic calculations (MuMax3~\cite{mumax}), using the parameters reported in Table~\ref{table:table1}\VJ{~\cite{note_simulation}}. In addition, an analytical model leading to very similar results has been developed~\cite{gehanne_in-preparation_2020}.
%
%
%
For the DW width, we use both the geometrical Hubert~\cite{hubert_book} $\Delta_H$ and dynamic Thiele~\cite{thiele_JAP_1974,thiaville_SDCMSIII_2006} $\Delta_T$ definitions. $\Delta_H$ corresponds to half of the distance separating the intersections between the slope of $M_z(x)$ at the domain wall center ($x=0$) and its value in the domains $M_z(x \rightarrow \pm \infty)$.
%
$\Delta_T$ is defined as $\Delta_T=2M_s^2/\int (d\overrightarrow{M}/dx)^2dx$~\cite{thiele_JAP_1974,thiaville_SDCMSIII_2006}. The predicted variations of $\Delta_H(H_x)$, $\Delta_T(H_x)$ and $\sigma(H_x)$ are reported in Fig.~\ref{fig:3} and follow the expected behaviors.
%
As the DMI field is perpendicular to the DW plane (i.e. $ \vec{H}_\mathrm{DMI} \parallel \vec{H}_x$), its contribution is essentially to shift the curves by $-H_\mathrm{DMI}$.
For $H_x$ varying from $-H_\mathrm{DMI}$, the magnetization within the DW progressively switches from Bloch ($\vec{M} \perp \vec{H}_x$) to Néel configuration ($\vec{M} \parallel \vec{H}_x$). 
As the DW width increases due to the Zeeman contribution ($\propto -\mu_0 \vec{M} \cdot (\vec{H}_x + \vec{H}_\mathrm{DMI}) $), the DW energy decreases.
%
Additionally, $\vec{H}_x$ tends to tilt the magnetization in the domains on both sides of the DW: $M_x(x \rightarrow \pm \infty)/M_s= H_x/H_{K_0}$. This reduces the angular variation of the magnetization within the DW and provokes a divergence of $\Delta_H(H_x)$ and $\Delta_T(H_x)$ and a vanishing $\sigma(H_x)$ for $H_x \rightarrow \pm H_{K_0}$. 


In order to discuss the DW pinning at the microscopic scale with in-plane field $H_x$, we use standard scaling arguments~\cite{lemerle_PRL_1998_domainwall_creep,agoritsas_PRE_13}. The free energy of a DW segment of length $L$, deformed over a distance $u$~\cite{lemerle_PRL_1998_domainwall_creep,hartmann_prb_2019} can be written as $\delta F(L,u)= \delta F_{elas}(L,u) - \delta F_{pin} (L,\xi)- \delta F_{z}(L,u)$,
where $\delta F_{elas}=\sigma t u^2 / L$ is the elastic energy produced by the increase of DW length and, $\delta F_{z}= 2\mu_0 M_s H t L u$ is the gain of Zeeman energy due to magnetization reversal. The DW stiffness $\sigma$ is, at this stage, assumed to be given by the DW surfacic energy.
For a weak disorder producing fluctuations of DW energy, the pinning term can be written $\delta F_{pin}= f_{pin}\sqrt{n \xi L}\xi$, where $f_{pin}$ and $\xi$ are the characteristic force and range of the DW-defects interaction, respectively and, $n$ is the density of pinning defects per unit surface area.
Here we go beyond the conventional relation $\xi \propto b$ ($\approx 1/\sqrt{n^2}$) used throughout in the literature~\cite{lemerle_PRL_1998_domainwall_creep}: we reintroduce a clear distinction between the characteristic length-scale of the pinning disorder $b$, which is fixed by the material inhomogeneities, and the range $\xi$ of the DW-defects interaction, which may vary with the DW structure~\cite{nattermann_prb_1990} as shown in the following discussion. 
For a short segment length ($L<L_c$), the DW is too rigid to follow the random pinning potential ($\delta F_{elas}(L,u)> \delta F_{pin}$). The segment is collectively pinned by a set of pinning defects. The characteristic collective pinning length-scale can be deduced from $\delta F_{elas}(L_c,\xi) \sim \delta F_{pin} (L_c,\xi)$, which reads $L_c \sim (\xi/n)^{1/3}(\sigma t/f_{pin})^{2/3}$. 
For $L>L_c$, the pinning energy is larger than the elastic energy: a DW can be seen as a set of rigid segments of length $L_c$, whose orientation follows the random pinning landscape. In order to depin a rigid segment, the magnetic field $H$ has to reach a threshold field $H_d'$. The latter can be determined from $\delta F_{pin}(L_c,\xi) \sim \delta F_{z} (L_c,\xi)$, which leads to $H_d' \sim \sqrt{n \xi/L_c} f_{pin}/(2\mu_0 M_st)$.
Then, assuming~\cite{jeudy_PRB_2018_DW_pinning} that the pinning barrier height is given by $k_BT_d \sim \delta F_{pin} (L_c,\xi)$, we obtain the scaling relations:   
\begin{eqnarray}
\label{eq: 3_xi}
\xi \sim \left[ (k_BT_d)^2/(2\mu_0 H_d'M_s\sigma t^2)\right]^{1/3},\\
\label{eq: 2_fpin}
f_{pin} \sim \VJ{\frac{1}{\sqrt{n}\xi}}\sqrt{2\mu_0 H_d'M_stk_BT_d},
\end{eqnarray}
which relate the characteristic range $\xi$ and force $f_{pin}$ of the DW-defect interaction, 
to the measured depinning field $H_d'(\mu_0 H_x)$ and temperature $T_d(\mu_0 H_x)$ (see Fig.~\ref{fig:1} d-e) and the predicted DW surface energy $\sigma(\mu_0 H_x)$ (see Fig.~\ref{fig:3} b).

The variation with in-plane field of $\xi$ and $f_{pin}$, determined from the measurements of $H_d'$, $T_d$ (see Fig.~\ref{fig:1} d-e), the predictions for $\sigma(\mu_0 H_x)$ (see Fig.~\ref{fig:3} b) and Eqs~\ref{eq: 3_xi}-~\ref{eq: 2_fpin}, are shown in Fig.~\ref{fig:3}.
%
For a comparison with the predicted variations of the DW width, the values of $\xi$ were rescaled with a constant factor, which is the only free parameter. 
As it can be observed in Fig.~\ref{fig:3} a, the variations with in-plane field of the pinning range $\xi(\mu_0 H_x)$ present strong correlations with the predictions for both $\Delta_H$ and $\Delta_T$: the agreement is very good for the Pt/Co/Au except for $\mu_0 H_x> 100~\mathrm{mT}$ and it is 
even better for the Pt/Co/Pt and Au/Co/Pt films. 
The DMI field seems to essentially modify the value of $\mu_0 H_x$  at which a minimum of DW width (and pinning range) is observed and no chiral contribution of pinning can be evidenced.
Those observations suggest that the range of the DW-defect interaction is close to the DW width.
%
As a direct consequence, the characteristic length-scale of pinning defects ($b \approx 1/\sqrt{n^2}$) should be close to or smaller than the DW width~\cite{nattermann_prb_1990}, which rules out the approximation considering DW as one dimensional line ~\cite{lemerle_PRL_1998_domainwall_creep,chauve_2000} interacting with remote defects ($\Delta_H$ and $\Delta_T \ll b$).

Let us now discuss the pinning force $f_{pin}$, whose variations with in-plane field is reported in Fig.\ref{fig:3} b. 
Here also, the only free parameter is the rescaling factor. As it can be observed, the pinning force follows rather well the variations of $\sigma(H_x)$. This is expected since the weak pinning of DWs results from fluctuations of DW energy, whose amplitude should decrease as $\sigma$ decreases. An additional contribution is the increase of DW width, which (for $b<\Delta$) is predicted to reduce the strength of pinning interaction~\cite{franken_jpcm_2012, soucaille_phd_2016}.

In conclusion, variations of the range and the strength of pinning following modifications of the wall structure have been evidenced. This effect, observed on magnetic domain walls with chiral structure submitted to large hard-axis magnetic fields, should be relevant for elastic interfaces moving in weak pinning disordered media~\cite{kolton_prb_2009_pathways,agoritsas_physicaB_12} in a wide variety of others systems.

We wish to thank J. Sampaio for careful reading of the manuscript.	

\bibliography{refs_creep_DMI}

\begin{thebibliography}{35}%
\makeatletter
\providecommand \@ifxundefined [1]{%
 \@ifx{#1\undefined}
}%
\providecommand \@ifnum [1]{%
 \ifnum #1\expandafter \@firstoftwo
 \else \expandafter \@secondoftwo
 \fi
}%
\providecommand \@ifx [1]{%
 \ifx #1\expandafter \@firstoftwo
 \else \expandafter \@secondoftwo
 \fi
}%
\providecommand \natexlab [1]{#1}%
\providecommand \enquote  [1]{``#1''}%
\providecommand \bibnamefont  [1]{#1}%
\providecommand \bibfnamefont [1]{#1}%
\providecommand \citenamefont [1]{#1}%
\providecommand \href@noop [0]{\@secondoftwo}%
\providecommand \href [0]{\begingroup \@sanitize@url \@href}%
\providecommand \@href[1]{\@@startlink{#1}\@@href}%
\providecommand \@@href[1]{\endgroup#1\@@endlink}%
\providecommand \@sanitize@url [0]{\catcode `\\12\catcode `\$12\catcode
  `\&12\catcode `\#12\catcode `\^12\catcode `\_12\catcode `\%12\relax}%
\providecommand \@@startlink[1]{}%
\providecommand \@@endlink[0]{}%
\providecommand \url  [0]{\begingroup\@sanitize@url \@url }%
\providecommand \@url [1]{\endgroup\@href {#1}{\urlprefix }}%
\providecommand \urlprefix  [0]{URL }%
\providecommand \Eprint [0]{\href }%
\providecommand \doibase [0]{http://dx.doi.org/}%
\providecommand \selectlanguage [0]{\@gobble}%
\providecommand \bibinfo  [0]{\@secondoftwo}%
\providecommand \bibfield  [0]{\@secondoftwo}%
\providecommand \translation [1]{[#1]}%
\providecommand \BibitemOpen [0]{}%
\providecommand \bibitemStop [0]{}%
\providecommand \bibitemNoStop [0]{.\EOS\space}%
\providecommand \EOS [0]{\spacefactor3000\relax}%
\providecommand \BibitemShut  [1]{\csname bibitem#1\endcsname}%
\let\auto@bib@innerbib\@empty
\bibitem [{\citenamefont {Thiaville}\ \emph {et~al.}(2012)\citenamefont
  {Thiaville}, \citenamefont {Rohart}, \citenamefont {Ju\'e}, \citenamefont
  {Cros},\ and\ \citenamefont {A.~Fert}}]{thiaville_EPL_2012}%
  \BibitemOpen
  \bibfield  {author} {\bibinfo {author} {\bibfnamefont {A.}~\bibnamefont
  {Thiaville}}, \bibinfo {author} {\bibfnamefont {S.}~\bibnamefont {Rohart}},
  \bibinfo {author} {\bibfnamefont {E.}~\bibnamefont {Ju\'e}}, \bibinfo
  {author} {\bibfnamefont {V.}~\bibnamefont {Cros}}, \ and\ \bibinfo {author}
  {\bibfnamefont {A.}~\bibnamefont {A.~Fert}},\ }\href {\doibase
  10.1209/0295-5075/100/57002} {\bibfield  {journal} {\bibinfo  {journal}
  {EPL}\ }\textbf {\bibinfo {volume} {100}},\ \bibinfo {pages} {57002}
  (\bibinfo {year} {2012})}\BibitemShut {NoStop}%
\bibitem [{\citenamefont {Caretta}\ \emph {et~al.}(2018)\citenamefont
  {Caretta}, \citenamefont {Mann}, \citenamefont {Büttner}, \citenamefont
  {Ueda}, \citenamefont {Pfau.}, \citenamefont {Günther}, \citenamefont
  {Hessing}, \citenamefont {Churikova}, \citenamefont {Klose}, \citenamefont
  {Schneider}, \citenamefont {Engel}, \citenamefont {Marcus}, \citenamefont
  {Bono}, \citenamefont {Bagschik}, \citenamefont {Eisebitt},\ and\
  \citenamefont {Beach}}]{caretta_natnano_2018}%
  \BibitemOpen
  \bibfield  {author} {\bibinfo {author} {\bibfnamefont {L.}~\bibnamefont
  {Caretta}}, \bibinfo {author} {\bibfnamefont {M.}~\bibnamefont {Mann}},
  \bibinfo {author} {\bibfnamefont {F.}~\bibnamefont {Büttner}}, \bibinfo
  {author} {\bibfnamefont {K.}~\bibnamefont {Ueda}}, \bibinfo {author}
  {\bibfnamefont {B.}~\bibnamefont {Pfau.}}, \bibinfo {author} {\bibfnamefont
  {C.}~\bibnamefont {Günther}}, \bibinfo {author} {\bibfnamefont
  {P.}~\bibnamefont {Hessing}}, \bibinfo {author} {\bibfnamefont
  {A.}~\bibnamefont {Churikova}}, \bibinfo {author} {\bibfnamefont
  {C.}~\bibnamefont {Klose}}, \bibinfo {author} {\bibfnamefont
  {M.}~\bibnamefont {Schneider}}, \bibinfo {author} {\bibfnamefont
  {D.}~\bibnamefont {Engel}}, \bibinfo {author} {\bibfnamefont
  {C.}~\bibnamefont {Marcus}}, \bibinfo {author} {\bibfnamefont
  {D.}~\bibnamefont {Bono}}, \bibinfo {author} {\bibfnamefont {K.}~\bibnamefont
  {Bagschik}}, \bibinfo {author} {\bibfnamefont {S.}~\bibnamefont {Eisebitt}},
  \ and\ \bibinfo {author} {\bibfnamefont {G.~S.~D.}\ \bibnamefont {Beach}},\
  }\href {\doibase 10.1038/s41565-018-0255-3} {\bibfield  {journal} {\bibinfo
  {journal} {Nature Nanotechnolog}\ }\textbf {\bibinfo {volume} {13}},\
  \bibinfo {pages} {1154} (\bibinfo {year} {2018})}\BibitemShut {NoStop}%
\bibitem [{\citenamefont {Fert}\ \emph {et~al.}(2013)\citenamefont {Fert},
  \citenamefont {Cros},\ and\ \citenamefont {Sampaio}}]{fert_natnano_2013}%
  \BibitemOpen
  \bibfield  {author} {\bibinfo {author} {\bibfnamefont {A.}~\bibnamefont
  {Fert}}, \bibinfo {author} {\bibfnamefont {V.}~\bibnamefont {Cros}}, \ and\
  \bibinfo {author} {\bibfnamefont {J.}~\bibnamefont {Sampaio}},\ }\href
  {\doibase http://dx.doi.org/10.1038/nnano.2013.29} {\bibfield  {journal}
  {\bibinfo  {journal} {Nature Nanotech}\ }\textbf {\bibinfo {volume} {8}},\
  \bibinfo {pages} {152–156} (\bibinfo {year} {2013})}\BibitemShut {NoStop}%
\bibitem [{\citenamefont {Ferrero}\ \emph {et~al.}(2017)\citenamefont
  {Ferrero}, \citenamefont {Foini}, \citenamefont {Giamarchi}, \citenamefont
  {Kolton},\ and\ \citenamefont
  {Rosso}}]{ferrero_prl_2017_spatiotemporal_patterns}%
  \BibitemOpen
  \bibfield  {author} {\bibinfo {author} {\bibfnamefont {E.~E.}\ \bibnamefont
  {Ferrero}}, \bibinfo {author} {\bibfnamefont {L.}~\bibnamefont {Foini}},
  \bibinfo {author} {\bibfnamefont {T.}~\bibnamefont {Giamarchi}}, \bibinfo
  {author} {\bibfnamefont {A.~B.}\ \bibnamefont {Kolton}}, \ and\ \bibinfo
  {author} {\bibfnamefont {A.}~\bibnamefont {Rosso}},\ }\href {\doibase
  10.1103/PhysRevLett.118.147208} {\bibfield  {journal} {\bibinfo  {journal}
  {Phys. Rev. Lett.}\ }\textbf {\bibinfo {volume} {118}},\ \bibinfo {pages}
  {147208} (\bibinfo {year} {2017})}\BibitemShut {NoStop}%
\bibitem [{\citenamefont {Grassi}\ \emph {et~al.}(2018)\citenamefont {Grassi},
  \citenamefont {Kolton}, \citenamefont {Jeudy}, \citenamefont {Mougin},
  \citenamefont {Bustingorry},\ and\ \citenamefont
  {Curiale}}]{grassi_prb_2018}%
  \BibitemOpen
  \bibfield  {author} {\bibinfo {author} {\bibfnamefont {M.~P.}\ \bibnamefont
  {Grassi}}, \bibinfo {author} {\bibfnamefont {A.~B.}\ \bibnamefont {Kolton}},
  \bibinfo {author} {\bibfnamefont {V.}~\bibnamefont {Jeudy}}, \bibinfo
  {author} {\bibfnamefont {A.}~\bibnamefont {Mougin}}, \bibinfo {author}
  {\bibfnamefont {S.}~\bibnamefont {Bustingorry}}, \ and\ \bibinfo {author}
  {\bibfnamefont {J.}~\bibnamefont {Curiale}},\ }\href {\doibase
  10.1103/PhysRevB.98.224201} {\bibfield  {journal} {\bibinfo  {journal} {Phys.
  Rev. B}\ }\textbf {\bibinfo {volume} {98}},\ \bibinfo {pages} {224201}
  (\bibinfo {year} {2018})}\BibitemShut {NoStop}%
\bibitem [{\citenamefont {Je}\ \emph {et~al.}(2013)\citenamefont {Je},
  \citenamefont {Kim}, \citenamefont {Yoo}, \citenamefont {Min}, \citenamefont
  {Lee},\ and\ \citenamefont {Choe}}]{je_prb_2013_DMI}%
  \BibitemOpen
  \bibfield  {author} {\bibinfo {author} {\bibfnamefont {S.-G.}\ \bibnamefont
  {Je}}, \bibinfo {author} {\bibfnamefont {D.-H.}\ \bibnamefont {Kim}},
  \bibinfo {author} {\bibfnamefont {S.-C.}\ \bibnamefont {Yoo}}, \bibinfo
  {author} {\bibfnamefont {B.-C.}\ \bibnamefont {Min}}, \bibinfo {author}
  {\bibfnamefont {K.-J.}\ \bibnamefont {Lee}}, \ and\ \bibinfo {author}
  {\bibfnamefont {S.-B.}\ \bibnamefont {Choe}},\ }\href {\doibase
  10.1103/PhysRevB.88.214401} {\bibfield  {journal} {\bibinfo  {journal} {Phys.
  Rev. B}\ }\textbf {\bibinfo {volume} {88}},\ \bibinfo {pages} {214401}
  (\bibinfo {year} {2013})}\BibitemShut {NoStop}%
\bibitem [{\citenamefont {Lavrijsen}\ \emph {et~al.}(2015)\citenamefont
  {Lavrijsen}, \citenamefont {Hartmann}, \citenamefont {van~den Brink},
  \citenamefont {Yin}, \citenamefont {Barcones}, \citenamefont {Duine},
  \citenamefont {Verheijen}, \citenamefont {Swagten},\ and\ \citenamefont
  {Koopmans}}]{lavrijsen_prb_2015}%
  \BibitemOpen
  \bibfield  {author} {\bibinfo {author} {\bibfnamefont {R.}~\bibnamefont
  {Lavrijsen}}, \bibinfo {author} {\bibfnamefont {D.~M.~F.}\ \bibnamefont
  {Hartmann}}, \bibinfo {author} {\bibfnamefont {A.}~\bibnamefont {van~den
  Brink}}, \bibinfo {author} {\bibfnamefont {Y.}~\bibnamefont {Yin}}, \bibinfo
  {author} {\bibfnamefont {B.}~\bibnamefont {Barcones}}, \bibinfo {author}
  {\bibfnamefont {R.~A.}\ \bibnamefont {Duine}}, \bibinfo {author}
  {\bibfnamefont {M.~A.}\ \bibnamefont {Verheijen}}, \bibinfo {author}
  {\bibfnamefont {H.~J.~M.}\ \bibnamefont {Swagten}}, \ and\ \bibinfo {author}
  {\bibfnamefont {B.}~\bibnamefont {Koopmans}},\ }\href {\doibase
  10.1103/PhysRevB.91.104414} {\bibfield  {journal} {\bibinfo  {journal} {Phys.
  Rev. B}\ }\textbf {\bibinfo {volume} {91}},\ \bibinfo {pages} {104414}
  (\bibinfo {year} {2015})}\BibitemShut {NoStop}%
\bibitem [{\citenamefont {Lau}\ \emph {et~al.}(2016)\citenamefont {Lau},
  \citenamefont {Sundar}, \citenamefont {Zhu},\ and\ \citenamefont
  {Sokalski}}]{lau_prb_2016}%
  \BibitemOpen
  \bibfield  {author} {\bibinfo {author} {\bibfnamefont {D.}~\bibnamefont
  {Lau}}, \bibinfo {author} {\bibfnamefont {V.}~\bibnamefont {Sundar}},
  \bibinfo {author} {\bibfnamefont {J.-G.}\ \bibnamefont {Zhu}}, \ and\
  \bibinfo {author} {\bibfnamefont {V.}~\bibnamefont {Sokalski}},\ }\href
  {\doibase 10.1103/PhysRevB.94.060401} {\bibfield  {journal} {\bibinfo
  {journal} {Phys. Rev. B}\ }\textbf {\bibinfo {volume} {94}},\ \bibinfo
  {pages} {060401} (\bibinfo {year} {2016})}\BibitemShut {NoStop}%
\bibitem [{\citenamefont {Pellegren}\ \emph {et~al.}(2017)\citenamefont
  {Pellegren}, \citenamefont {Lau},\ and\ \citenamefont
  {Sokalski}}]{pellegrin_prl_2017}%
  \BibitemOpen
  \bibfield  {author} {\bibinfo {author} {\bibfnamefont {J.~P.}\ \bibnamefont
  {Pellegren}}, \bibinfo {author} {\bibfnamefont {D.}~\bibnamefont {Lau}}, \
  and\ \bibinfo {author} {\bibfnamefont {V.}~\bibnamefont {Sokalski}},\ }\href
  {\doibase 10.1103/PhysRevLett.119.027203} {\bibfield  {journal} {\bibinfo
  {journal} {Phys. Rev. Lett.}\ }\textbf {\bibinfo {volume} {119}},\ \bibinfo
  {pages} {027203} (\bibinfo {year} {2017})}\BibitemShut {NoStop}%
\bibitem [{\citenamefont {Lau}\ \emph {et~al.}(2018)\citenamefont {Lau},
  \citenamefont {Pellegren}, \citenamefont {Nembach}, \citenamefont {Shaw},\
  and\ \citenamefont {Sokalski}}]{lau_prb_2018}%
  \BibitemOpen
  \bibfield  {author} {\bibinfo {author} {\bibfnamefont {D.}~\bibnamefont
  {Lau}}, \bibinfo {author} {\bibfnamefont {J.~P.}\ \bibnamefont {Pellegren}},
  \bibinfo {author} {\bibfnamefont {H.~T.}\ \bibnamefont {Nembach}}, \bibinfo
  {author} {\bibfnamefont {J.~M.}\ \bibnamefont {Shaw}}, \ and\ \bibinfo
  {author} {\bibfnamefont {V.}~\bibnamefont {Sokalski}},\ }\href {\doibase
  10.1103/PhysRevB.98.184410} {\bibfield  {journal} {\bibinfo  {journal} {Phys.
  Rev. B}\ }\textbf {\bibinfo {volume} {98}},\ \bibinfo {pages} {184410}
  (\bibinfo {year} {2018})}\BibitemShut {NoStop}%
\bibitem [{\citenamefont {Hartmann}\ \emph {et~al.}(2019)\citenamefont
  {Hartmann}, \citenamefont {Duine}, \citenamefont {Meijer}, \citenamefont
  {Swagten},\ and\ \citenamefont {Lavrijsen}}]{hartmann_prb_2019}%
  \BibitemOpen
  \bibfield  {author} {\bibinfo {author} {\bibfnamefont {D.~M.~F.}\
  \bibnamefont {Hartmann}}, \bibinfo {author} {\bibfnamefont {R.~A.}\
  \bibnamefont {Duine}}, \bibinfo {author} {\bibfnamefont {M.~J.}\ \bibnamefont
  {Meijer}}, \bibinfo {author} {\bibfnamefont {H.~J.~M.}\ \bibnamefont
  {Swagten}}, \ and\ \bibinfo {author} {\bibfnamefont {R.}~\bibnamefont
  {Lavrijsen}},\ }\href {\doibase 10.1103/PhysRevB.100.094417} {\bibfield
  {journal} {\bibinfo  {journal} {Phys. Rev. B}\ }\textbf {\bibinfo {volume}
  {100}},\ \bibinfo {pages} {094417} (\bibinfo {year} {2019})}\BibitemShut
  {NoStop}%
\bibitem [{\citenamefont {Lemerle}\ \emph {et~al.}(1998)\citenamefont
  {Lemerle}, \citenamefont {Ferr{\'e}}, \citenamefont {Chappert}, \citenamefont
  {Mathet}, \citenamefont {Giamarchi},\ and\ \citenamefont {{Le
  Doussal}}}]{lemerle_PRL_1998_domainwall_creep}%
  \BibitemOpen
  \bibfield  {author} {\bibinfo {author} {\bibfnamefont {S.}~\bibnamefont
  {Lemerle}}, \bibinfo {author} {\bibfnamefont {J.}~\bibnamefont {Ferr{\'e}}},
  \bibinfo {author} {\bibfnamefont {C.}~\bibnamefont {Chappert}}, \bibinfo
  {author} {\bibfnamefont {V.}~\bibnamefont {Mathet}}, \bibinfo {author}
  {\bibfnamefont {T.}~\bibnamefont {Giamarchi}}, \ and\ \bibinfo {author}
  {\bibfnamefont {P.}~\bibnamefont {{Le Doussal}}},\ }\href@noop {} {\bibfield
  {journal} {\bibinfo  {journal} {Phys. Rev. Lett.}\ }\textbf {\bibinfo
  {volume} {80}},\ \bibinfo {pages} {849} (\bibinfo {year} {1998})}\BibitemShut
  {NoStop}%
\bibitem [{\citenamefont {Jeudy}\ \emph {et~al.}(2016)\citenamefont {Jeudy},
  \citenamefont {Mougin}, \citenamefont {Bustingorry}, \citenamefont
  {Savero~Torres}, \citenamefont {Gorchon}, \citenamefont {Kolton},
  \citenamefont {Lema\^{\i}tre},\ and\ \citenamefont
  {Jamet}}]{jeudy_PRL_2016_energy_barrier}%
  \BibitemOpen
  \bibfield  {author} {\bibinfo {author} {\bibfnamefont {V.}~\bibnamefont
  {Jeudy}}, \bibinfo {author} {\bibfnamefont {A.}~\bibnamefont {Mougin}},
  \bibinfo {author} {\bibfnamefont {S.}~\bibnamefont {Bustingorry}}, \bibinfo
  {author} {\bibfnamefont {W.}~\bibnamefont {Savero~Torres}}, \bibinfo {author}
  {\bibfnamefont {J.}~\bibnamefont {Gorchon}}, \bibinfo {author} {\bibfnamefont
  {A.~B.}\ \bibnamefont {Kolton}}, \bibinfo {author} {\bibfnamefont
  {A.}~\bibnamefont {Lema\^{\i}tre}}, \ and\ \bibinfo {author} {\bibfnamefont
  {J.-P.}\ \bibnamefont {Jamet}},\ }\href {\doibase
  10.1103/PhysRevLett.117.057201} {\bibfield  {journal} {\bibinfo  {journal}
  {Phys. Rev. Lett.}\ }\textbf {\bibinfo {volume} {117}},\ \bibinfo {pages}
  {057201} (\bibinfo {year} {2016})}\BibitemShut {NoStop}%
\bibitem [{\citenamefont {Diaz~Pardo}\ \emph {et~al.}(2017)\citenamefont
  {Diaz~Pardo}, \citenamefont {Savero~Torres}, \citenamefont {Kolton},
  \citenamefont {Bustingorry},\ and\ \citenamefont
  {Jeudy}}]{diaz_PRB_2017_depinning}%
  \BibitemOpen
  \bibfield  {author} {\bibinfo {author} {\bibfnamefont {R.}~\bibnamefont
  {Diaz~Pardo}}, \bibinfo {author} {\bibfnamefont {W.}~\bibnamefont
  {Savero~Torres}}, \bibinfo {author} {\bibfnamefont {A.~B.}\ \bibnamefont
  {Kolton}}, \bibinfo {author} {\bibfnamefont {S.}~\bibnamefont {Bustingorry}},
  \ and\ \bibinfo {author} {\bibfnamefont {V.}~\bibnamefont {Jeudy}},\ }\href
  {\doibase 10.1103/PhysRevB.95.184434} {\bibfield  {journal} {\bibinfo
  {journal} {Phys. Rev. B}\ }\textbf {\bibinfo {volume} {95}},\ \bibinfo
  {pages} {184434} (\bibinfo {year} {2017})}\BibitemShut {NoStop}%
\bibitem [{\citenamefont {Edwards}\ and\ \citenamefont
  {Wilkinson}(1982)}]{edwards_wilkinson_1982}%
  \BibitemOpen
  \bibfield  {author} {\bibinfo {author} {\bibfnamefont {S.~F.}\ \bibnamefont
  {Edwards}}\ and\ \bibinfo {author} {\bibfnamefont {D.~R.}\ \bibnamefont
  {Wilkinson}},\ }\href {http://www.jstor.org/stable/2397363} {\bibfield
  {journal} {\bibinfo  {journal} {Proceedings of the Royal Society of London.
  Series A, Mathematical and Physical Sciences}\ }\textbf {\bibinfo {volume}
  {381}},\ \bibinfo {pages} {17} (\bibinfo {year} {1982})}\BibitemShut
  {NoStop}%
\bibitem [{\citenamefont {Kolton}\ \emph {et~al.}(2009)\citenamefont {Kolton},
  \citenamefont {Rosso}, \citenamefont {Giamarchi},\ and\ \citenamefont
  {Krauth}}]{kolton_prb_2009_pathways}%
  \BibitemOpen
  \bibfield  {author} {\bibinfo {author} {\bibfnamefont {A.~B.}\ \bibnamefont
  {Kolton}}, \bibinfo {author} {\bibfnamefont {A.}~\bibnamefont {Rosso}},
  \bibinfo {author} {\bibfnamefont {T.}~\bibnamefont {Giamarchi}}, \ and\
  \bibinfo {author} {\bibfnamefont {W.}~\bibnamefont {Krauth}},\ }\href@noop {}
  {\bibfield  {journal} {\bibinfo  {journal} {Phys. Rev. B}\ }\textbf {\bibinfo
  {volume} {79}},\ \bibinfo {pages} {184207} (\bibinfo {year}
  {2009})}\BibitemShut {NoStop}%
\bibitem [{\citenamefont {Agoritsas}\ \emph {et~al.}(2012)\citenamefont
  {Agoritsas}, \citenamefont {Lecomte},\ and\ \citenamefont
  {Giamarchi}}]{agoritsas_physicaB_12}%
  \BibitemOpen
  \bibfield  {author} {\bibinfo {author} {\bibfnamefont {E.}~\bibnamefont
  {Agoritsas}}, \bibinfo {author} {\bibfnamefont {V.}~\bibnamefont {Lecomte}},
  \ and\ \bibinfo {author} {\bibfnamefont {T.}~\bibnamefont {Giamarchi}},\
  }\href {\doibase https://doi.org/10.1016/j.physb.2012.01.017} {\bibfield
  {journal} {\bibinfo  {journal} {Physica B: Condensed Matter}\ }\textbf
  {\bibinfo {volume} {407}},\ \bibinfo {pages} {1725} (\bibinfo {year}
  {2012})},\ \bibinfo {note} {proceedings of the International Workshop on
  Electronic Crystals (ECRYS-2011)}\BibitemShut {NoStop}%
\bibitem [{\citenamefont {Jeudy}\ \emph {et~al.}(2018)\citenamefont {Jeudy},
  \citenamefont {D\'{\i}az~Pardo}, \citenamefont {Savero~Torres}, \citenamefont
  {Bustingorry},\ and\ \citenamefont {Kolton}}]{jeudy_PRB_2018_DW_pinning}%
  \BibitemOpen
  \bibfield  {author} {\bibinfo {author} {\bibfnamefont {V.}~\bibnamefont
  {Jeudy}}, \bibinfo {author} {\bibfnamefont {R.}~\bibnamefont
  {D\'{\i}az~Pardo}}, \bibinfo {author} {\bibfnamefont {W.}~\bibnamefont
  {Savero~Torres}}, \bibinfo {author} {\bibfnamefont {S.}~\bibnamefont
  {Bustingorry}}, \ and\ \bibinfo {author} {\bibfnamefont {A.~B.}\ \bibnamefont
  {Kolton}},\ }\href {\doibase 10.1103/PhysRevB.98.054406} {\bibfield
  {journal} {\bibinfo  {journal} {Phys. Rev. B}\ }\textbf {\bibinfo {volume}
  {98}},\ \bibinfo {pages} {054406} (\bibinfo {year} {2018})}\BibitemShut
  {NoStop}%
\bibitem [{\citenamefont {Agoritsas}\ \emph {et~al.}(2013)\citenamefont
  {Agoritsas}, \citenamefont {Lecomte},\ and\ \citenamefont
  {Giamarchi}}]{agoritsas_PRE_13}%
  \BibitemOpen
  \bibfield  {author} {\bibinfo {author} {\bibfnamefont {E.}~\bibnamefont
  {Agoritsas}}, \bibinfo {author} {\bibfnamefont {V.}~\bibnamefont {Lecomte}},
  \ and\ \bibinfo {author} {\bibfnamefont {T.}~\bibnamefont {Giamarchi}},\
  }\href {\doibase 10.1103/PhysRevE.87.042406} {\bibfield  {journal} {\bibinfo
  {journal} {Phys. Rev. E}\ }\textbf {\bibinfo {volume} {87}},\ \bibinfo
  {pages} {042406} (\bibinfo {year} {2013})}\BibitemShut {NoStop}%
\bibitem [{\citenamefont {Gross}\ \emph {et~al.}(2018)\citenamefont {Gross},
  \citenamefont {Akhtar}, \citenamefont {Hrabec}, \citenamefont {Sampaio},
  \citenamefont {Mart\'{\i}nez}, \citenamefont {Chouaieb}, \citenamefont
  {Shields}, \citenamefont {Maletinsky}, \citenamefont {Thiaville},
  \citenamefont {Rohart},\ and\ \citenamefont {Jacques}}]{gross_PRM_2018}%
  \BibitemOpen
  \bibfield  {author} {\bibinfo {author} {\bibfnamefont {I.}~\bibnamefont
  {Gross}}, \bibinfo {author} {\bibfnamefont {W.}~\bibnamefont {Akhtar}},
  \bibinfo {author} {\bibfnamefont {A.}~\bibnamefont {Hrabec}}, \bibinfo
  {author} {\bibfnamefont {J.}~\bibnamefont {Sampaio}}, \bibinfo {author}
  {\bibfnamefont {L.~J.}\ \bibnamefont {Mart\'{\i}nez}}, \bibinfo {author}
  {\bibfnamefont {S.}~\bibnamefont {Chouaieb}}, \bibinfo {author}
  {\bibfnamefont {B.~J.}\ \bibnamefont {Shields}}, \bibinfo {author}
  {\bibfnamefont {P.}~\bibnamefont {Maletinsky}}, \bibinfo {author}
  {\bibfnamefont {A.}~\bibnamefont {Thiaville}}, \bibinfo {author}
  {\bibfnamefont {S.}~\bibnamefont {Rohart}}, \ and\ \bibinfo {author}
  {\bibfnamefont {V.}~\bibnamefont {Jacques}},\ }\href@noop {} {\bibfield
  {journal} {\bibinfo  {journal} {Phys. Rev. Mat.}\ }\textbf {\bibinfo {volume}
  {2}},\ \bibinfo {pages} {024406} (\bibinfo {year} {2018})}\BibitemShut
  {NoStop}%
\bibitem [{\citenamefont {Shahbazi}\ \emph {et~al.}(2018)\citenamefont
  {Shahbazi}, \citenamefont {Hrabec}, \citenamefont {Moretti}, \citenamefont
  {Ward}, \citenamefont {Moore}, \citenamefont {Jeudy}, \citenamefont
  {Martinez},\ and\ \citenamefont {Marrows}}]{shahbazi_PRB_2018}%
  \BibitemOpen
  \bibfield  {author} {\bibinfo {author} {\bibfnamefont {K.}~\bibnamefont
  {Shahbazi}}, \bibinfo {author} {\bibfnamefont {A.}~\bibnamefont {Hrabec}},
  \bibinfo {author} {\bibfnamefont {S.}~\bibnamefont {Moretti}}, \bibinfo
  {author} {\bibfnamefont {M.~B.}\ \bibnamefont {Ward}}, \bibinfo {author}
  {\bibfnamefont {T.~A.}\ \bibnamefont {Moore}}, \bibinfo {author}
  {\bibfnamefont {V.}~\bibnamefont {Jeudy}}, \bibinfo {author} {\bibfnamefont
  {E.}~\bibnamefont {Martinez}}, \ and\ \bibinfo {author} {\bibfnamefont
  {C.~H.}\ \bibnamefont {Marrows}},\ }\href {\doibase
  10.1103/PhysRevB.98.214413} {\bibfield  {journal} {\bibinfo  {journal} {Phys.
  Rev. B}\ }\textbf {\bibinfo {volume} {98}},\ \bibinfo {pages} {214413}
  (\bibinfo {year} {2018})}\BibitemShut {NoStop}%
\bibitem [{\citenamefont {Shahbazi}\ \emph {et~al.}(2019)\citenamefont
  {Shahbazi}, \citenamefont {Kim}, \citenamefont {Nembach}, \citenamefont
  {Shaw}, \citenamefont {Bischof}, \citenamefont {Rossell}, \citenamefont
  {Jeudy}, \citenamefont {Moore},\ and\ \citenamefont
  {Marrows}}]{shahbazi_PRB_2019}%
  \BibitemOpen
  \bibfield  {author} {\bibinfo {author} {\bibfnamefont {K.}~\bibnamefont
  {Shahbazi}}, \bibinfo {author} {\bibfnamefont {J.-V.}\ \bibnamefont {Kim}},
  \bibinfo {author} {\bibfnamefont {H.~T.}\ \bibnamefont {Nembach}}, \bibinfo
  {author} {\bibfnamefont {J.~M.}\ \bibnamefont {Shaw}}, \bibinfo {author}
  {\bibfnamefont {A.}~\bibnamefont {Bischof}}, \bibinfo {author} {\bibfnamefont
  {M.~D.}\ \bibnamefont {Rossell}}, \bibinfo {author} {\bibfnamefont
  {V.}~\bibnamefont {Jeudy}}, \bibinfo {author} {\bibfnamefont {T.~A.}\
  \bibnamefont {Moore}}, \ and\ \bibinfo {author} {\bibfnamefont {C.~H.}\
  \bibnamefont {Marrows}},\ }\href {\doibase 10.1103/PhysRevB.99.094409}
  {\bibfield  {journal} {\bibinfo  {journal} {Phys. Rev. B}\ }\textbf {\bibinfo
  {volume} {99}},\ \bibinfo {pages} {094409} (\bibinfo {year}
  {2019})}\BibitemShut {NoStop}%
\bibitem [{\citenamefont {Bustingorry}\ \emph {et~al.}(2012)\citenamefont
  {Bustingorry}, \citenamefont {Kolton},\ and\ \citenamefont
  {Giamarchi}}]{bustingorry_PRE_12_thermal_rounding}%
  \BibitemOpen
  \bibfield  {author} {\bibinfo {author} {\bibfnamefont {S.}~\bibnamefont
  {Bustingorry}}, \bibinfo {author} {\bibfnamefont {A.~B.}\ \bibnamefont
  {Kolton}}, \ and\ \bibinfo {author} {\bibfnamefont {T.}~\bibnamefont
  {Giamarchi}},\ }\href {\doibase 10.1103/PhysRevE.85.021144} {\bibfield
  {journal} {\bibinfo  {journal} {Phys. Rev. E}\ }\textbf {\bibinfo {volume}
  {85}},\ \bibinfo {pages} {021144} (\bibinfo {year} {2012})}\BibitemShut
  {NoStop}%
\bibitem [{\citenamefont {Ju\'e}\ \emph {et~al.}(2016)\citenamefont {Ju\'e},
  \citenamefont {Thiaville}, \citenamefont {Pizzini}, \citenamefont {Miltat},
  \citenamefont {Sampaio}, \citenamefont {Buda-Prejbeanu}, \citenamefont
  {Rohart}, \citenamefont {Vogel}, \citenamefont {Bonfim}, \citenamefont
  {Boulle}, \citenamefont {Auffret}, \citenamefont {Miron},\ and\ \citenamefont
  {Gaudin}}]{Jue_PRB_2016}%
  \BibitemOpen
  \bibfield  {author} {\bibinfo {author} {\bibfnamefont {E.}~\bibnamefont
  {Ju\'e}}, \bibinfo {author} {\bibfnamefont {A.}~\bibnamefont {Thiaville}},
  \bibinfo {author} {\bibfnamefont {S.}~\bibnamefont {Pizzini}}, \bibinfo
  {author} {\bibfnamefont {J.}~\bibnamefont {Miltat}}, \bibinfo {author}
  {\bibfnamefont {J.}~\bibnamefont {Sampaio}}, \bibinfo {author} {\bibfnamefont
  {L.~D.}\ \bibnamefont {Buda-Prejbeanu}}, \bibinfo {author} {\bibfnamefont
  {S.}~\bibnamefont {Rohart}}, \bibinfo {author} {\bibfnamefont
  {J.}~\bibnamefont {Vogel}}, \bibinfo {author} {\bibfnamefont
  {M.}~\bibnamefont {Bonfim}}, \bibinfo {author} {\bibfnamefont
  {O.}~\bibnamefont {Boulle}}, \bibinfo {author} {\bibfnamefont
  {S.}~\bibnamefont {Auffret}}, \bibinfo {author} {\bibfnamefont {I.~M.}\
  \bibnamefont {Miron}}, \ and\ \bibinfo {author} {\bibfnamefont
  {G.}~\bibnamefont {Gaudin}},\ }\href {\doibase 10.1103/PhysRevB.93.014403}
  {\bibfield  {journal} {\bibinfo  {journal} {Phys. Rev. B}\ }\textbf {\bibinfo
  {volume} {93}},\ \bibinfo {pages} {014403} (\bibinfo {year}
  {2016})}\BibitemShut {NoStop}%
\bibitem [{not({\natexlab{a}})}]{note_Hd_effectif}%
  \BibitemOpen
  \href@noop {} {\emph {\bibinfo {title} {As $H_x$ remains significantly
  smaller than $H_{K_0}$, the contribution of the magnetization tilting is
  small: the difference between uncorrected and corrected values of $\xi$ is
  less than the size of the data points of Fig. 2a, i.e., much less than the
  error bars.}}}\BibitemShut {Stop}%
\bibitem [{\citenamefont {Vansteenkiste}\ \emph {et~al.}(2014)\citenamefont
  {Vansteenkiste}, \citenamefont {Leliaert}, \citenamefont {Dvornik},
  \citenamefont {Helsen}, \citenamefont {Garcia-Sanchez},\ and\ \citenamefont
  {{Van Waeyenberge}}}]{mumax}%
  \BibitemOpen
  \bibfield  {author} {\bibinfo {author} {\bibfnamefont {A.}~\bibnamefont
  {Vansteenkiste}}, \bibinfo {author} {\bibfnamefont {J.}~\bibnamefont
  {Leliaert}}, \bibinfo {author} {\bibfnamefont {M.}~\bibnamefont {Dvornik}},
  \bibinfo {author} {\bibfnamefont {M.}~\bibnamefont {Helsen}}, \bibinfo
  {author} {\bibfnamefont {F.}~\bibnamefont {Garcia-Sanchez}}, \ and\ \bibinfo
  {author} {\bibfnamefont {B.}~\bibnamefont {{Van Waeyenberge}}},\ }\href
  {\doibase 10.1063/1.4899186} {\bibfield  {journal} {\bibinfo  {journal} {AIP
  Advances}\ }\textbf {\bibinfo {volume} {4}},\ \bibinfo {pages} {107133}
  (\bibinfo {year} {2014})}\BibitemShut {NoStop}%
\bibitem [{not({\natexlab{b}})}]{note_simulation}%
  \BibitemOpen
  \href@noop {} {\emph {\bibinfo {title} {The numerical calculations were
  performed on $1024\times1\times1$ cubic cells, with 0.9 nm edge size.
  Periodic boundary conditions were enforced in the $y$ direction, the
  magnetostatic interaction coefficients being computed over 100 000
  repetitions on each side.}}}\BibitemShut {Stop}%
\bibitem [{\citenamefont {G\'ehanne}\ \emph {et~al.}()\citenamefont
  {G\'ehanne}, \citenamefont {Thiaville}, \citenamefont {Rohart},\ and\
  \citenamefont {Jeudy}}]{gehanne_in-preparation_2020}%
  \BibitemOpen
  \bibfield  {author} {\bibinfo {author} {\bibfnamefont {P.}~\bibnamefont
  {G\'ehanne}}, \bibinfo {author} {\bibfnamefont {A.}~\bibnamefont
  {Thiaville}}, \bibinfo {author} {\bibfnamefont {S.}~\bibnamefont {Rohart}}, \
  and\ \bibinfo {author} {\bibfnamefont {V.}~\bibnamefont {Jeudy}},\
  }\href@noop {} {\bibinfo  {journal} {in preparation}\ }\BibitemShut {NoStop}%
\bibitem [{\citenamefont {Hubert}(1974)}]{hubert_book}%
  \BibitemOpen
\bibfield  {journal} {  }\bibfield  {author} {\bibinfo {author} {\bibfnamefont
  {A.}~\bibnamefont {Hubert}},\ }\href@noop {} {\emph {\bibinfo {title}
  {Theorie der Domänenwände in Geordneten Medien (in german)}}}\ (\bibinfo
  {publisher} {Springer},\ \bibinfo {year} {1974})\BibitemShut {NoStop}%
\bibitem [{\citenamefont {Thiele}(1974)}]{thiele_JAP_1974}%
  \BibitemOpen
  \bibfield  {author} {\bibinfo {author} {\bibfnamefont {A.~A.}\ \bibnamefont
  {Thiele}},\ }\href {\doibase 10.1063/1.1662989} {\bibfield  {journal}
  {\bibinfo  {journal} {Journal of Applied Physics}\ }\textbf {\bibinfo
  {volume} {45}},\ \bibinfo {pages} {377} (\bibinfo {year} {1974})},\ \Eprint
  {http://arxiv.org/abs/https://doi.org/10.1063/1.1662989}
  {https://doi.org/10.1063/1.1662989} \BibitemShut {NoStop}%
\bibitem [{\citenamefont {Thiaville}\ and\ \citenamefont
  {Nakatani}(2006)}]{thiaville_SDCMSIII_2006}%
  \BibitemOpen
  \bibfield  {author} {\bibinfo {author} {\bibfnamefont {A.}~\bibnamefont
  {Thiaville}}\ and\ \bibinfo {author} {\bibfnamefont {Y.}~\bibnamefont
  {Nakatani}},\ }\enquote {\bibinfo {title} {Spin dynamics in confined magnetic
  structures iii},}\ \ (\bibinfo  {publisher} {Springer},\ \bibinfo {address}
  {Berlin},\ \bibinfo {year} {2006})\ Chap.\ \bibinfo {chapter} {Domain wall
  dynamics in nanowires and nanostrips}, pp.\ \bibinfo {pages}
  {161--206}\BibitemShut {NoStop}%
\bibitem [{\citenamefont {Nattermann}\ \emph {et~al.}(1990)\citenamefont
  {Nattermann}, \citenamefont {Shapir},\ and\ \citenamefont
  {Vilfan}}]{nattermann_prb_1990}%
  \BibitemOpen
  \bibfield  {author} {\bibinfo {author} {\bibfnamefont {T.}~\bibnamefont
  {Nattermann}}, \bibinfo {author} {\bibfnamefont {Y.}~\bibnamefont {Shapir}},
  \ and\ \bibinfo {author} {\bibfnamefont {I.}~\bibnamefont {Vilfan}},\ }\href
  {\doibase 10.1103/PhysRevB.42.8577} {\bibfield  {journal} {\bibinfo
  {journal} {Phys. Rev. B}\ }\textbf {\bibinfo {volume} {42}},\ \bibinfo
  {pages} {8577} (\bibinfo {year} {1990})}\BibitemShut {NoStop}%
\bibitem [{\citenamefont {Chauve}\ \emph {et~al.}(2000)\citenamefont {Chauve},
  \citenamefont {Giamarchi},\ and\ \citenamefont {Le~Doussal}}]{chauve_2000}%
  \BibitemOpen
  \bibfield  {author} {\bibinfo {author} {\bibfnamefont {P.}~\bibnamefont
  {Chauve}}, \bibinfo {author} {\bibfnamefont {T.}~\bibnamefont {Giamarchi}}, \
  and\ \bibinfo {author} {\bibfnamefont {P.}~\bibnamefont {Le~Doussal}},\
  }\href {\doibase 10.1103/PhysRevB.62.6241} {\bibfield  {journal} {\bibinfo
  {journal} {Phys. Rev. B}\ }\textbf {\bibinfo {volume} {62}},\ \bibinfo
  {pages} {6241} (\bibinfo {year} {2000})}\BibitemShut {NoStop}%
\bibitem [{\citenamefont {Franken}\ \emph {et~al.}(2012)\citenamefont
  {Franken}, \citenamefont {Hoeijmakers}, \citenamefont {Lavrijsen},\ and\
  \citenamefont {Swagten}}]{franken_jpcm_2012}%
  \BibitemOpen
  \bibfield  {author} {\bibinfo {author} {\bibfnamefont {J.~H.}\ \bibnamefont
  {Franken}}, \bibinfo {author} {\bibfnamefont {M.}~\bibnamefont
  {Hoeijmakers}}, \bibinfo {author} {\bibfnamefont {R.}~\bibnamefont
  {Lavrijsen}}, \ and\ \bibinfo {author} {\bibfnamefont {H.~J.~M.}\
  \bibnamefont {Swagten}},\ }\href
  {http://stacks.iop.org/0953-8984/24/i=2/a=024216} {\bibfield  {journal}
  {\bibinfo  {journal} {Journal of Physics: Condensed Matter}\ }\textbf
  {\bibinfo {volume} {24}},\ \bibinfo {pages} {024216} (\bibinfo {year}
  {2012})}\BibitemShut {NoStop}%
\bibitem [{\citenamefont {Soucaille}(2016)}]{soucaille_phd_2016}%
  \BibitemOpen
  \bibfield  {author} {\bibinfo {author} {\bibfnamefont {R.}~\bibnamefont
  {Soucaille}},\ }\emph {\bibinfo {title} {De l'importance de l'interaction de
  Dzyaloshinskii-Moriya sur la dynamique sous champ des parois de domaines
  magn{\'e}tiques dans des films d{\'e}sordonn{\'e}s (in french)
  https://tel.archives-ouvertes.fr/tel-01495704}},\ \href
  {https://tel.archives-ouvertes.fr/tel-01495704} {\bibinfo {type} {Theses}},\
  \bibinfo  {school} {{Universit{\'e} Paris-Saclay}} (\bibinfo {year} {2016}),\
  \Eprint {http://arxiv.org/abs/https://tel.archives-ouvertes.fr/tel-01495704}
  {https://tel.archives-ouvertes.fr/tel-01495704} \BibitemShut {NoStop}%
\end{thebibliography}%

\end{document}